\documentclass[11pt, a4paper]{article}

\usepackage[utf8]{inputenc}
\usepackage[T1]{fontenc}
\usepackage{geometry}
\usepackage{tikz}
\usetikzlibrary{shapes, arrows.meta, positioning, calc, backgrounds, shadows}

\definecolor{TechBlue}{RGB}{0, 85, 150}
\definecolor{SoftTeal}{RGB}{80, 180, 180}
\definecolor{Crimson}{RGB}{180, 40, 40}
\usepackage{titlesec}
\usepackage{parskip}      
\usepackage{amsmath}
\usepackage{amssymb}
\usepackage{booktabs}     
\usepackage{enumitem}     
\usepackage{wrapfig}      
\usepackage{fancyhdr}     
\usepackage[numbers,sort&compress]{natbib} 
\usepackage{hyperref}     
\usepackage{cleveref}

\usepackage{tgheros}

\definecolor{DeepNavy}{RGB}{20, 40, 70}
\definecolor{AccentBlue}{RGB}{0, 85, 150}
\definecolor{LightGray}{RGB}{240, 240, 240}

\titleformat{\section}
  {\Large\bfseries\color{DeepNavy}}
  {\thesection}{1em}{}
  
\titleformat{\subsection}
  {\large\bfseries\color{AccentBlue}}
  {\thesubsection}{1em}{}

\hypersetup{
    colorlinks=true,
    linkcolor=DeepNavy,
    urlcolor=AccentBlue,
    citecolor=AccentBlue,
    pdftitle={TMM Market Study}
}

\begin{document}

\begin{center}
    \vspace*{1cm}
    {\Huge \textbf{\color{DeepNavy} The Transfer Matrix Method (TMM)}} \\
    \vspace{0.5cm}
    {\Large \textit{\color{black} Technical Bottlenecks and Industrial Evolution in 1D Wave Physics}} \\
    \vspace{0.5cm}

    \textbf{Author:} Joaquin Garc\'ia Su\'arez \
    \textbf{Date:} \today \\
    \vspace{1cm}
    \hrule
\end{center}
\vspace{1cm}

\begin{abstract}
\noindent The Transfer Matrix Method (TMM) stands as the ubiquitous computational backbone for analyzing 1D wave propagation in layered media, underpinning critical product designs in photonics, seismology, and acoustics—industries collectively valued in the tens of USD billions. 
Despite its essential role, legacy implementations of TMM create significant technical (and therefore strategic) bottlenecks, primarily due to a lack of straightforward differentiability and high computational costs associated with Uncertainty Quantification (UQ). 
This white paper assesses the current market footprint of TMM, identifies the economic ``hidden costs'' of traditional workflows, and outlines an emerging industrial alternative -- Differentiable Programming and Neural Surrogates -- and their own limitations.
\end{abstract}

\section*{Executive Summary}
The Transfer Matrix Method (TMM) is a widely used \emph{forward} model for wave propagation through stratified media and a core computational workhorse for thin-film optical coating design and related layered-wave applications \cite{Mackay:2021,Jimenez:2021}.
Since early formulations in mid-20th-century optics \cite{Abeles:1950}, the core formalism has remained stable, and as a result it is deeply embedded in industrial toolchains.

These tool chains sit inside large commercial markets: optical coatings alone are commonly estimated in the \(\sim\)USD 16--22B range for 2024 (depending on scope), while the broader photonics ecosystem is reported at hundreds of billions (and photonics-enabled products at much larger scale) \cite{GrandView:2024OptCoat,FBI:2024OptCoat,SPIE:2025Strength,Spectaris:2025Trend}.
As inverse design becomes mainstream across optics and adjacent domains, engineering teams increasingly require fast gradients and scalable optimization loops \cite{Molesky:2018,Morrison:2024TMMAdjoint}.

In practice, the bottleneck is rarely the mathematics of TMM, but the \emph{workflow}: many legacy implementations do not expose gradients or integrate cleanly with modern automatic-differentiation (AD) stacks.
This pushes teams toward finite differences or derivative-free heuristics and makes tolerance/yield analysis disproportionately expensive \cite{Baydin:2018,Danis:2025TMMax, Li:2025JaxLayerLumos}.

This document quantifies the economic ``hidden tax'' of such legacy workflows (compute spend, engineering time, and iteration latency), and surveys two mitigation families: (i) differentiable TMM/adjoint-enabled formulations, and (ii) surrogate models that preserve physical constraints while accelerating large-scale search and uncertainty studies \cite{Morrison:2024TMMAdjoint,Danis:2025TMMax, Li:2025JaxLayerLumos}.

\section{The Industrial Backbone: Where TMM is Critical}
In 1D stratified media, the Transfer Matrix Method (TMM) computes reflection/transmission by (i) expressing fields in each homogeneous layer as a superposition of forward/backward eigenmodes and (ii) enforcing boundary continuity, yielding a matrix product that maps incident to transmitted states (see \Cref{fig:tmm_schematic}) \cite{Yeh:1988,Katsidis:2002}.
This same computational primitive reappears across industries under different names (e.g., the Thomson--Haskell propagator matrix in seismology) \cite{thomson1950transmission,Haskell:1953}.

\begin{figure}[h]
\centering
\begin{tikzpicture}[font=\sffamily\small]
    \definecolor{DeepNavy}{RGB}{20, 40, 70}
    \definecolor{AccentBlue}{RGB}{0, 85, 150}
    
    \tikzset{
        layer/.style={draw=none, minimum width=3cm, minimum height=4cm},
        arrow/.style={->, >=stealth, thick, DeepNavy},
        dim/.style={|<->|, gray, thin}
    }

    \fill[gray!5] (-3,0) rectangle (0,4);
    \node at (-1.5, 3.8) {\textbf{Ambient} ($n_0$)};
    
    \fill[AccentBlue!10] (0,0) rectangle (2,4);
    \node at (1, 0.3) {Layer 1};
    \draw[dim] (0,-0.2) -- (2,-0.2) node[midway, below] {$d_1$};
    
    \fill[AccentBlue!25] (2,0) rectangle (3.5,4);
    \node at (2.75, 0.3) {Layer 2};
    \draw[dim] (2,-0.2) -- (3.5,-0.2) node[midway, below] {$d_2$};
    
    \node at (4.25, 2) {$\dots$};
    
    \fill[DeepNavy!10] (5,0) rectangle (8,4);
    \node at (6.5, 3.8) {\textbf{Substrate} ($n_s$)};

    \draw[gray, thin] (0,0) -- (0,4);
    \draw[gray, thin] (2,0) -- (2,4);
    \draw[gray, thin] (3.5,0) -- (3.5,4);
    \draw[gray, thin] (5,0) -- (5,4);

    \draw[arrow] (-2.5, 3.5) -- (0, 2.5) node[midway, above, sloped] {$E_{in}$};
    \draw[arrow] (0, 2.5) -- (-2.5, 1.5) node[midway, below, sloped] {$E_{r}$};
    
    \draw[DeepNavy, dashed] (0, 2.5) -- (2, 2.0) -- (3.5, 2.3) -- (5, 1.8);
    
    \draw[arrow] (5, 1.8) -- (7.5, 1.2) node[midway, above, sloped] {$E_{t}$};


\end{tikzpicture}

\caption{\textbf{Schematic of the Transfer Matrix Method.} The transfer matrix relates the fields at a given cross-section of the medium to the fields anywhere else via the multiplication of characteristic matrices for each discrete layer, i.e., the effect of each layer is condensed in one matrix, the overall medium is characterized by the ordered product of the matrices that make it up. 
Unlike transfer matrices, scattering matrices connect right-going (e.g., $E_{in}$, $E_t$) and left-going amplitudes ($E_{r}$) at different locations within the stack. }
\label{fig:tmm_schematic}
\end{figure}

\subsection{Photonics \& Optics (Dominant Commercial Footprint)}
In optical thin films, TMM is the workhorse that links target spectra (e.g., $R(\lambda,\theta)$, phase, group delay) to multilayer stack designs \cite{Macleod:2017,Katsidis:2002,Mitsas:1995}.
\begin{itemize}[leftmargin=*]
    \item \textbf{Applications:} Anti-reflective (AR) coatings, dielectric mirrors/DBRs, interference filters, VCSEL cavity mirrors, photonic-crystal-like 1D stacks.
    \item \textbf{Impact:} Core to high-volume optical coating supply chains and also to high-stakes multilayer reflective optics in EUV/BEUV lithography \cite{Schroder:2007,Uzoma:2021}.
    \item \textbf{Technical Usage:} Engineers choose materials and layer thicknesses to minimize a merit function over wavelength/angle/polarization, often with tolerance constraints; this drives large batches of forward evaluations in optimization loops \cite{WildBuhay:1986,Lininger:2021}.
\end{itemize}

\subsection{Seismology \& Geophysics}
In this domain the method is commonly introduced as the \textit{Thomson--Haskell} or \textit{propagator matrix} formalism \cite{thomson1950transmission,Haskell:1953,Aki-Richards}.
\begin{itemize}[leftmargin=*]
    \item \textbf{Applications:} 1D site response and layered-earth modeling (e.g., shear-wave amplification from bedrock to surface).
    \item \textbf{Impact:} Used in seismic-hazard workflows for compliance in seismic zones and in energy exploration/near-surface stratigraphy characterization.
    \item \textbf{Technical Usage:} Propagates state vectors (e.g., displacement--stress) through layers to predict amplification/transfer functions that can be inverted against observations in site identification problems \cite{thomson1950transmission,Haskell:1953}.
\end{itemize}

\subsection{Phononics \& Acoustics}
\begin{itemize}[leftmargin=*]
    \item \textbf{Applications:} Acoustic metamaterial panels, layered liners/impedance tubes, sonar domes, ultrasonic matching layers, 1D phononic crystals.
    \item \textbf{Impact:} Enables engineered bandgaps and reflection bands (``sound mirrors'') for vibration isolation and noise control \cite{Hussein:2014,Laude:2020}.
    \item \textbf{Technical Usage:} Combined with periodicity/Bloch conditions, transfer matrices yield dispersion relations and bandgaps for layered/periodic acoustic structures \cite{Dai:2024}.
\end{itemize}

\section{Strategic Bottlenecks: The ``Hidden Tax'' of Legacy TMM} 
TMM is computationally efficient for the \textit{forward} problem (predicting performance of a given stack), but inverse design turns that speed into limited business value when gradients, yield analysis, and numerical conditioning are missing from the workflows.

\subsection{The Gradient Gap (The Optimization Tax)}
Standard TMM is \textit{differentiable} in principle; the practical bottleneck is that those gradients are extremely costly to compute (involve many extra matrix chain multiplications), and teams therefore default to finite-difference or derivative-free searches.
\begin{itemize}[leftmargin=*]
    \item \textbf{The Bottleneck:} Gradients are often unavailable in production toolchains (``TMM as a black box''), not because the mathematics forbids them, but because the code does.
    \item \textbf{Consequence:} Reliance on zeroth-order methods (Genetic Algorithms, Particle Swarm) or finite-difference gradients that scale poorly with parameter count, inflating compute spend and slowing iteration cycles.
    \item \textbf{Trend (What changes the economics):} Adjoint methods and automatic differentiation (``differentiable programming'') make gradients first-class, enabling gradient-based optimization directly on layer thicknesses and material parameters \cite{Hughes:2018,Molesky:2018,Morrison:2024TMMAdjoint,Li:2025JaxLayerLumos,Danis:2025TMMax}.
\end{itemize}

\begin{figure}[h]
\centering
\begin{tikzpicture}[
    font=\sffamily\small,
    node distance=2.5cm,
    block/.style={rectangle, draw=DeepNavy, thick, fill=white, 
                  text width=2.5cm, align=center, rounded corners, minimum height=1.2cm},
    arrow/.style={->, >=stealth, thick, DeepNavy},
    gradarrow/.style={->, >=stealth, thick, dashed, red!70!black}
    ]
    
    \definecolor{DeepNavy}{RGB}{20, 40, 70}
    
    \node[block] (params) {\textbf{Design Parameters}\\($x_1, x_2, \dots$)};
    
    \node[block, right of=params, xshift=2.5cm, fill=AccentBlue!10] (tmm) 
        {\textbf{Differentiable TMM}\\(PyTorch/JAX)};
        
    \node[block, right of=tmm, xshift=2.5cm] (loss) 
        {\textbf{Loss Function}\\$\mathcal{L}=|Target - Current|$};
        
    \draw[arrow] (params) -- node[above] {Forward} (tmm);
    \draw[arrow] (tmm) -- node[above] {Spectrum} (loss);
    
    \draw[gradarrow] (loss.south) -- ++(0,-0.8) -| node[pos=0.25, below] 
        {\textbf{Automatic Differentiation} ($\nabla_{\mathbf{x}}\mathcal{L}$)} (params.south);

    \node[anchor=north, text=gray, align=center, yshift=-1.5cm] at (tmm.south) 
        {\textit{Replaces legacy ("blind") global searches}};

\end{tikzpicture}
\caption{\textbf{The Inverse Design Loop Enabled by Automatic Differentiation} By making the TMM differentiable, gradients (dashed red line) propagate automatically from the Loss function back to the Design Parameters, enabling more efficient optimization than ``global methods'' such as GAs (genetic algorithms) or particle swarm methods.}
\label{fig:diff_tmm}
\end{figure}


\begin{figure}[h]
\centering
\begin{tikzpicture}[font=\sffamily\footnotesize]

    \draw[thick, <->] (0, 4) node[left] {Loss ($\mathcal{L}$)} -- (0, 0) -- (11, 0) node[below] {Design Parameter Space};
    
    \draw[thick, DeepNavy] plot [smooth, tension=0.7] coordinates {
        (0.2, 3.5) (2, 2.5) (4, 3.2) (7, 0.5) (9, 2.0) (10.5, 3.5)
    };
    \node[DeepNavy, above] at (7, 0.0) {Local Optimum};

    
    \foreach \x/\y in {0.5/3.4, 1.0/3.2, 1.5/2.8, 2.5/2.6, 9.5/2.5, 10/3.0} {
        \fill[gray!50] (\x, \y) circle (2pt);
    }
    \foreach \x/\y in {3.5/3.0, 4.5/2.8, 5.5/1.5, 8.5/1.8} {
        \fill[gray!80] (\x, \y) circle (2pt);
    }
    \foreach \x/\y in {6.5/0.8, 7.5/0.7, 6.8/0.6} {
        \fill[DeepNavy] (\x, \y) circle (2pt);
    }
    
    \node[anchor=west, text=gray] at (0.5, 3.8) {\textbf{A. Genetic Algorithms (Legacy)}};
    \draw[->, gray, thin] (2, 3.6) -- (1, 3.3); 
    \node[text=gray, font=\scriptsize, align=left] at (2.2, 4.5) {Random "population" sampling\\Requires $10^4+$ simulations};

    
    \fill[Crimson] (3.8, 3.25) circle (2.5pt) node[above] {Start};
    
    \draw[->, >=Stealth, thick, Crimson, dashed] (3.8, 3.25) -- (5.5, 1.8) -- (6.5, 0.8) -- (6.9, 0.55);
    
    \node[anchor=east, text=Crimson] at (10.5, 3.8) {\textbf{B. Gradient Descent (AI-Ready)}};
    \node[text=Crimson, font=\scriptsize, align=right] at (8, 3.0) {Follows the slope ($\nabla \mathcal{L}$)\\Converges in $\sim 100$ steps};

    
    \node[draw=none, fill=LightGray!50, rounded corners, minimum width=10cm, minimum height=1.8cm] at (5.5, -1.5) {};
    
    \node[anchor=west] at (0.5, -0.8) {\textbf{Computational Cost (Time to Solution)}};
    
    \node[anchor=east, font=\scriptsize] at (2.5, -1.3) {GA+TMM (legacy)};
    \filldraw[fill=gray!70, draw=none] (2.6, -1.45) rectangle (9.5, -1.15);
    \node[anchor=west, font=\scriptsize, text=gray] at (9.6, -1.3) {$\sim$ hours (CPU Cluster)};
    
    \node[anchor=east, font=\scriptsize] at (2.5, -1.8) {Differentiable TMM};
    \filldraw[fill=Crimson, draw=none] (2.6, -1.95) rectangle (3.2, -1.65); 
    \node[anchor=west, font=\scriptsize, text=Crimson] at (3.3, -1.8) {$\sim$ minutes (CPU or GPU)};

\end{tikzpicture}
\caption{\textbf{The Efficiency Gap:} While legacy methods (A) rely on stochastic "trial, keep most apt, evolve and try again", Differentiable TMM (B) utilizes numerical exact gradients to reach the optimal design orders of magnitude faster and with less computational effort.}
\label{fig:optimization_gap}
\end{figure}

\subsection{Uncertainty Quantification (UQ) and Yield Costs}
Fabrication tolerances (e.g., thickness variations of $\pm 1\%$) force \textit{yield} thinking: what fraction of manufactured parts still meet spec?
\begin{itemize}[leftmargin=*]
    \item \textbf{The Bottleneck:} Robustness estimation typically requires Monte Carlo (or surrogate-assisted) evaluation across many perturbed stacks; complexity grows further when deposition error compensation or in-situ re-optimization is modeled \cite{Koss:1992,Tikhonravov:2011,Poitras:2025,Sullivan:1992,Dai:2022Yield}.
    \item \textbf{Consequence:} When rigorous yield analysis is too expensive, organizations substitute ``safety factors'' (over-engineering) or broaden specs, increasing material cost and reducing achievable performance.
\end{itemize}

\subsection{Numerical Conditioning (When TMM Becomes Fragile)}
In the presence of evanescent waves, strong absorption, or very thick stacks, transfer-matrix products can become ill-conditioned (overflow/underflow), degrading accuracy.
\begin{itemize}[leftmargin=*]
    \item \textbf{Consequence:} Teams migrate to numerically stable formulations (scattering-matrix / S-matrix propagation) that trade simplicity for robustness \cite{Cotter:1995,Auslender:1996SMatrix,Bay:2022PyLlama}.
\end{itemize}

\section{Market Evolution and Solutions} 
The pivot is not from ``Classic TMM'' to ``ML+TMM'', 
instead, teams are modernizing \emph{workflows around TMM} to (i) expose gradients, (ii) accelerate yield/robustness loops, and (iii) eliminate numerical fragility in extreme stacks.
In practice, the winning pattern is often \textbf{hybrid}: differentiable forward models for optimization, surrogates for high-throughput screening/UQ, and stable propagators where conditioning breaks.

\begin{table}[h]
\centering
\renewcommand{\arraystretch}{1.35}
\begin{tabular}{p{3.0cm} p{3.7cm} p{6.2cm}@{}}
\toprule
\textbf{Pain Point} & \textbf{Emerging Solution} & \textbf{Technical Implementation} \\ \midrule
Gradients missing in incumbent workflows 
& \textbf{Differentiable / Adjoint TMM} 
& AD-ready re-implementations (e.g., JAX/PyTorch) or adjoint formulations that make $\nabla_{\mathbf{x}}\mathrm{Loss}$ first-class, enabling scalable gradient-based inverse design \cite{Baydin:2018,Morrison:2024TMMAdjoint,Li:2025JaxLayerLumos,Danis:2025TMMax}. \\
\hline
Expensive yield / robustness loops (UQ) 
& \textbf{Surrogates + Robust Design} 
& Neural or surrogate models trained on TMM evaluations to accelerate screening and Monte Carlo; paired with yield-aware/robust merit functions and (ideally) calibrated uncertainty estimates \cite{peurifoy2018nanophotonic,Lininger:2021,Dai:2022Yield,Poitras:2025}. \\
\hline
Numerical ill-conditioning for thick/evanescent regimes 
& \textbf{Stable Propagation (S-Matrix)} 
& Replace transfer-matrix products with scattering-matrix (S-matrix) propagation (or related stabilized formulations) to avoid overflow/underflow and preserve accuracy in extreme stacks \cite{Cotter:1995,Auslender:1996SMatrix,Bay:2022PyLlama}. \\
\bottomrule
\end{tabular}
\end{table}

\section{Economic Impact Assessment}

\begin{figure}[t]
\centering
\includegraphics[width=0.62\linewidth]{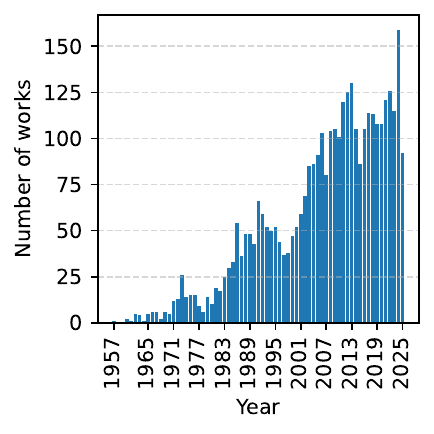}
\caption{From OpenAlex \cite{OpenAlex:API}: evolution of the number of academic works with titles matching ``transfer matrix method'' (query via the OpenAlex \texttt{works} endpoint using \texttt{filter=title.search} and grouped by \texttt{publication\_year}). Year 2024 is the highest in this query (\(\gtrsim 150\) items). 2025 counts are partial-year (as of June in our pull). This intentionally undercounts papers that use TMM but do not mention it in the title.}
\label{fig:trend}
\end{figure}

The economic footprint of TMM is indirect but substantial: it is embedded in design, verification, and yield-analysis loops across layered-wave technologies. Rather than treating TMM as a standalone ``market,'' we anchor impact through (i) end-market scale and (ii) the software/tooling budgets that monetize faster optimization, yield, and robustness.

\begin{itemize}[leftmargin=*]
    \item \textbf{Optical coatings (direct TMM footprint) + precision optics (context):}
    Optical coatings are commonly estimated at \(\sim\)USD 16.7B (2024) \(\rightarrow\) USD 30.6B (2032) \cite{FBI:2024OptCoat} or \(\sim\)USD 22.1B (2024) \(\rightarrow\) USD 37.6B (2030) \cite{GrandView:2024OptCoat}, depending on scope.
    Precision optics is separately reported at \(\sim\)USD 27.3B (2024) \(\rightarrow\) USD 40.0B (2030) \cite{RMI:2025PrecisionOptics}.
    In these workflows, differentiable/adjoint TMM and yield-aware optimization primarily monetize through reduced iteration latency and improved pass-yield (see also \cite{Tikhonravov:2011,Sullivan:1992,Dai:2022Yield}).

    \item \textbf{Seismic services and interpretation software (propagator-matrix workflows):}
    Seismic services are commonly placed around USD 8.6--9.0B in 2024 \cite{IMARC:2024Seismic,GMI:2024Seismic}.
    Within that, market estimates for seismic interpretation software vary, e.g., \(\sim\)USD 2.18B (2024) \(\rightarrow\) USD 4.77B (2033) \cite{GMR:2024SeisInterp} with other reports giving higher baselines \cite{DataIntelo:2024SeisInterp}.
    The monetization lever is similar: faster inversion loops, and more rigorous uncertainty handling at acceptable cost.

    \item \textbf{Acoustics / NVH (captured under CAE and simulation software spend):}
    Rather than quote brittle ``acoustics engineering'' TAMs, we treat layered acoustic TMM use as part of broader simulation/CAE purchasing.
    The simulation software market is estimated at \(\sim\)USD 13.44B (2024) \cite{FBI:2024SimSoftware}, and the CAE market at \(\sim\)USD 8.91B (2024) \cite{FBI:2024CAE}.
    In this bucket, the opportunity is accelerating repeated evaluations (robust design / Monte Carlo / multi-objective loops) via differentiable solvers and surrogates.

    \item \textbf{Academia (innovation signal, not a primary revenue driver):}
    TMM remains a high-usage research tool and an active methods topic (Fig.~\ref{fig:trend}), but academic licensing revenue is typically minor relative to industrial design and manufacturing toolchains.
\end{itemize}

\noindent\textbf{Sizing note (transparent, assumption-driven):}
To avoid opaque ratios (e.g., ``5\% of TAM''), we treat TMM-enabled tooling as a niche within simulation/CAE spend and report ranges via sensitivity.
For example, if optics/photonics accounts for 2--6\% of simulation software spend and layered-media (coatings/1D stacks) accounts for 20--40\% of that optics slice, the implied envelope is:
\[
\text{TMM-enabled optics tooling} \approx (13.44\text{B}) \times (0.02\text{--}0.06) \times (0.20\text{--}0.40)
\approx 54\text{M--}323\text{M per year},
\]
with the wide range reflecting scope and procurement heterogeneity rather than false precision \cite{FBI:2024SimSoftware}.

\section{Perspectives: The Road Ahead}
The Transfer Matrix Method (TMM) is not disappearing; it remains a core \emph{forward} solver for 1D layered-wave physics.
What is evolving is how TMM can be embedded into industrial pipelines: gradients, robustness, and stability are becoming first-class requirements rather than afterthoughts.

\subsection{The ``AI'' Angle}
The practical shift is from forward-only simulation to \textbf{optimization- and UQ-ready} workflows built around (i) automatic differentiation/adjoints and (ii) surrogate models.
\begin{itemize}[leftmargin=*]
    \item \textbf{First-order inverse design:} Differentiable/adjoint-enabled TMM exposes $\nabla_{\mathbf{x}}\mathrm{Loss}$ for thicknesses/material parameters, enabling scalable gradient-based optimization (still non-convex, but far more sample-efficient than zeroth-order search) \cite{Baydin:2018,Molesky:2018,Morrison:2024TMMAdjoint,Li:2025JaxLayerLumos,Danis:2025TMMax}.
    \item \textbf{High-throughput screening and faster robustness loops:} Neural surrogates can approximate TMM mappings to accelerate large parameter sweeps and Monte Carlo-style yield studies, provided uncertainty is calibrated and the domain of validity is respected \cite{peurifoy2018nanophotonic,Lininger:2021,Poitras:2025,Dai:2022Yield}.
\end{itemize}

\subsection{Reality Check: Technical Limitations}
Despite clear upside, ``differentiable TMM'' and surrogates face predictable engineering constraints.
\begin{itemize}[leftmargin=*]
    \item \textbf{AutoDiff overhead (memory \& runtime):} Backpropagation through long stacks and broadband objectives can increase memory and compute requirements.
    Practical mitigations include gradient checkpointing and stabilized propagation choices \cite{Griewank:2008,Chen:2016Checkpointing,Bay:2022PyLlama}.
    
    \item \textbf{Distribution shift and calibration risk:} Surrogates can fail under dataset shift (new materials, thickness regimes, angles/polarizations) unless coverage, uncertainty calibration, and validation gates are explicit \cite{Ovadia:2019Shift,Guo:2017Calibration}.
    
    \item \textbf{Cold-start cost:} High-fidelity surrogates require an upfront dataset (often generated by legacy solvers). For small or highly bespoke projects, the data-generation cost can dominate, making ``surrogate-first'' unattractive.
\end{itemize}

\subsection{Strategic Outlook}
Near term, the winning product pattern is \textbf{hybrid}:
a stable forward solver (TMM/S-matrix) remains in the loop, numerical gradients are exposed for design iterations, and surrogates are used selectively for throughput (screening, sensitivity, and UQ acceleration) \cite{Morrison:2024TMMAdjoint,Poitras:2025,Bay:2022PyLlama}.
This improves cycle time and yield engineering without sacrificing physical correctness; however, it does not automatically provide interpretability: by the end of the day, teams get numbers for the figures of merit but not insights as to what physical phenomenon within the layering is setting performance.
%


\bibliographystyle{unsrtnat}

\vspace{1cm}
\hrule
\vspace{0.3cm}
\footnotesize{
Reach J.G.S. at 
\texttt{joaquin.garciasuarez@epfl.ch} (academic inquiries)
or at 
\texttt{ajgarciasuarez@gmail.com} (others)
}

\end{document}